\documentclass[aps,prl,twocolumn,groupedaddress,showpacs,amsfonts,amssymb]{revtex4}

\usepackage{graphics}
\usepackage{epsfig}
\usepackage{amsmath}
\usepackage{amsthm}

\begin{document}

\title{Markovian feedback to control continuous variable entanglement}

\author{Stefano Mancini}
\email{stefano.mancini@unicam.it}

\affiliation{Dipartimento di Fisica, Universit\`{a} di Camerino, 
I-62032 Camerino, Italy\\
and INFN, Sezione di Perugia, I-06123 Perugia, Italy}

\date{\today}

\begin{abstract}
We present a model to realize quantum feedback control of
continuous variable entanglement.
It consists of two interacting bosonic modes subject to amplitude 
damping 
and achieving entangled Gaussian steady state. 
The possibility to greatly improve the degree of entanglement
by means of Markovian (direct) feedback is then shown. 
\end{abstract}


\maketitle

Entanglement  is one of the most puzzling features of  quantum 
mechanics that has given rise to long debates on foundational aspects 
since the seminal papers on the subject \cite{sch,epr}.
Recently, it has been also recognized as a valuable resource for 
quantum information processing \cite{bdv}. Thus, its generation and 
control has became a primary task to accomplish.
So far much attention has been devoted to generation and 
characterization of entanglement, while for its control one usually 
refers to error correction or distillation procedures \cite{bdv}.

Feedback provides an intuitive way to control a system state, hence 
entanglement since it is a system state peculiarity.
A theory of quantum-limited feedback has been introduced by Wiseman 
and Milburn \cite{wismil93} leading to relevant experimental 
achievements \cite{mab}.
It is based on the direct (immediate) use of the measurement results 
to alter the system state,
hence the name \textit{Markovian feedback}.
It turns out to be particularly useful in realizing squeezed states
\cite{wismil94}.
In fact, monitoring a system's observable will 
conditionally squeeze its variance. 
Unconditional squeezing is then obtained by using the measurement 
results to continuously drive the system into the desired, 
deterministic, squeezed state.
If such arguments are applied to a multiparties system in case of 
\textit{nonlocal} measurements, 
the realized squeezing is related to entanglement among the parties 
\cite{thom}.

\begin{figure}
\begin{center}
\includegraphics[width=0.4\textwidth]{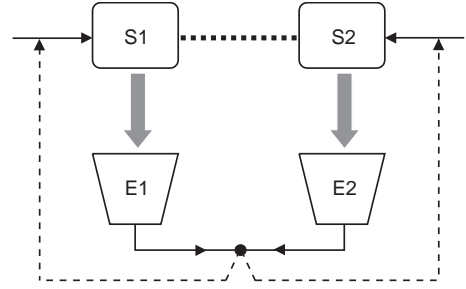}
\end{center}
\caption{\label{fig1} The studied scheme. Two subsystems $S1$ and 
$S2$ interact (thik dashed line).
Each one is coherently driven (left-right arrow and right-left arrow) 
and loses information in its own environment, $E1$ and $E2$. The 
environments are separately monitored and then a joint current is 
exploited for a feedback action on the driving fields (dashed lines).}
\end{figure}

However, the crucial point is to see whether \textit{local} 
measurements 
followed by feedback action suffice to control entanglement
(because local measurements condition the system to separable states).
We shall show that it is indeed possible. By referring to 
Fig.\ref{fig1}, we can consider two interacting subsystems $S1$ and 
$S2$
each one loosing information on its own environment $E1$ and $E2$. 
Then, monitoring such environments separately will give outcomes' 
currents that can be joined and exploited for feedback action.

Recent results on finite dimensional systems show that such feedback 
scheme is 
helpful in enforcing entanglement \cite{mw}. 
However, the original gedanken experiment proposal of Ref.\cite{epr} 
refers to infinite dimensional systems; moreover continuous variables 
offer a lot of advantages in information processing \cite{bp}.
We henceforth focus our attention on the possibility to control the 
degree of entanglement for a two-party quantum Gaussian state. 
A paradigm of the model of Ref.\cite{epr} is represented by a 
non-degenerate parametric oscillator \cite{reid} where two bosonic 
modes
$a_1$ and $a_2$ interact through a quadratic Hamiltonian of the type 
($\hbar=1$)
\begin{equation}\label{Hint}
H_{int}=i\chi\left(a_1^{\dag}a_2^{\dag}-a_1a_2\right)\,,
\end{equation}
with $\chi\in\mathbb{R}$ the coupling strength.
In addition we consider the two modes driven by a Hamiltonian
\begin{equation}\label{Hdrive}
H_{drive}=\alpha\left(a_1^{\dag}+a_2^{\dag}\right)+\alpha^*
\left(a_1+a_2\right)\,,
\end{equation}
where $\alpha\in\mathbb{C}$ is the driving amplitude (assumed equal 
for both modes).
Hence, the total Hamiltonian results
\begin{equation}\label{Htot}
H_{tot}=H_{int}+H_{drive}\,.
\end{equation}

Now, suppose that the system looses information on each mode through 
an amplitude damping channel at rate $\kappa$ (hereafter we consider 
$\kappa=1$ so that all physical quantities become dimensionless).
Then, by introducing the modes quadratures 
\begin{equation}\label{quad}
X_j=\frac{a_j+a_j^{\dag}}{\sqrt{2}}\,,
\quad
Y_j=\frac{a_j-a_j^{\dag}}{i\sqrt{2}}\,,
\quad
j=1,2\,,
\end{equation}
the system dynamics can easily be described by the 
quantum Langevin equations \cite{cg}
\begin{eqnarray}
\dot X_1&=&\chi 
X_2-i\frac{\alpha-\alpha^*}{\sqrt{2}}-\frac{1}{2}X_1+{\cal X}_1\,,
\label{lang1}
\\
\dot Y_1&=&-\chi 
Y_2-\frac{\alpha+\alpha^*}{\sqrt{2}}-\frac{1}{2}Y_1+{\cal Y}_1\,,
\label{lang2}
\\
\dot X_2&=&\chi 
X_1-i\frac{\alpha-\alpha^*}{\sqrt{2}}-\frac{1}{2}X_2+{\cal X}_2\,,
\label{lang3}
\\
\dot Y_2&=&-\chi 
Y_1-\frac{\alpha+\alpha^*}{\sqrt{2}}-\frac{1}{2}Y_2+{\cal Y}_2\,,
\label{lang4}
\end{eqnarray}
where ${\cal X}_j$ and ${\cal Y}_j$ ($j=1,2$) are hermitian 
operators representing the vacuum noise entering the system. They 
have equal time correlations
\begin{eqnarray}
\langle {\cal X}_j{\cal X}_k \rangle&=&
\langle {\cal Y}_j{\cal Y}_k \rangle=\frac{1}{2}\delta_{jk}\,,
\label{XXcorr}\\
\langle {\cal X}_j{\cal Y}_k \rangle&=&
-\langle {\cal Y}_j{\cal X}_k \rangle=i\frac{1}{2}\delta_{jk}\,,
\quad j,k=1,2\,.
\label{XYcorr}
\end{eqnarray}
By monitoring each mode environment it would be possible to measure a 
quadrature for each subsystem, e.g. by using homodyne detection 
\cite{wm93}.
Suppose this is done with an overall efficiency $\eta\in[0,1]$
(i.e., $\eta$ accounts for the detectors efficiency, the fraction of 
the field being measured, 
etc.), then we can write local currents like \cite{cg}
\begin{eqnarray}\label{Itj}
I_j(t)=\eta X_{j}+\sqrt{\eta}{\cal W}_j\,,\quad j=1,2\,,
\end{eqnarray}
where ${\cal W}_j$ ($j=1,2$) are hermitian operators representing the 
vacuum noise affecting the currents. They have equal 
time correlations
\begin{equation}\label{Wcorr}
\langle {\cal W}_j{\cal W}_k \rangle=\frac{1}{2}\delta_{jk}\,,
\quad j,k=1,2\,.
\end{equation}
We can now combine the currents $I_j$ to get the following joint 
current 
\begin{eqnarray}\label{It}
I(t)=\eta\left(X_{1}-X_{2}\right)+\sqrt{\eta}\left({\cal 
W}_1-{\cal W}_2\right)\,,
\end{eqnarray}
through which it would be possible to gain information about 
the quantity $X_1-X_2$. Its variance in absence of environmental 
effects and for increasing value of $\chi$, tends to zero  
\cite{reid},  corresponding to the maximally entangled state of the 
type discussed in Ref.\cite{epr}. Thus, the fact that 
it goes below 1 (vacuum fluctuations) can be roughly considered as 
signature of entanglement.
In such a sense, we may assume the current in Eq.(\ref{It}) giving us 
information about the system state entanglement.

The steady state solution of Eqs.(\ref{lang1})-(\ref{lang4}) results 
\begin{eqnarray}
\langle X_1 \rangle_{ss}&=&\langle 
X_2\rangle_{ss}=-i\sqrt{2}\;\frac{\alpha-\alpha^*}{1-2\chi}\,,
\label{Xss}\\
\langle Y_1 \rangle_{ss}&=&\langle 
Y_2\rangle_{ss}=-\sqrt{2}\;\frac{\alpha+\alpha^*}{1+2\chi}\,,
\label{Yss}
\end{eqnarray}
which is stable for $\chi<1/2$.

Subtracting Eq.(\ref{lang3}) from (\ref{lang1}) we get a quantum 
Langevin equation
resembling a Ornstein-Uhlenbeck process \cite{HSM}. 
It can be easily solved to get the steady state variance
\begin{equation}\label{var1-2}
\langle \left(X_1-X_2\right)^2\rangle-\left(\langle 
X_1\rangle_{ss}-\langle X_2\rangle_{ss}\right)^2=\frac{1}{1+2\chi}\,.
\end{equation}
We see that the variance (\ref{var1-2}) goes below $1$ (vacuum 
fluctuations) as soon as $\chi>0$, thus we infer the presence of 
entanglement in  the steady state. However, since it must be 
$\chi<1/2$, the variance
(\ref{var1-2}) is limited from below by $1/2$. By converse, the 
amount of entanglement will be limited. That is, the damping channel 
degrades the system state preventing it to become maximally entangled 
like that of Ref.\cite{epr}. 

We can now think to control such a process, hence to control 
entanglement,
by using feedback action  
accordingly to the information gained about $X_1-X_2$.
To this end we consider an additional Hamiltonian proportional to the 
current (\ref{It}) and driving the operator $Y_1-Y_2$, 
\begin{equation}\label{Hfb}
H_{fb}=\frac{\lambda}{\eta}\;I(t-\tau)
\left(Y_1-Y_2\right)\,.
\end{equation}
Here $\lambda$ represents the feedback strength and $\tau$ the 
feedback loop delay time.
The above choice of the driving is motivated by the fact that 
we want to affect the variance of 
$X_1-X_2$. The feedback action (\ref{Hfb}) can be realized by a 
modulation of the driving fields 
accordingly to the current $I$.

Adding the Hamiltonian (\ref{Hfb}) to the Langevin equations 
(\ref{lang1})-(\ref{lang4}) means to add, for a generic operator 
${\cal O}$, the term 
\begin{equation}\label{Ofb}
\dot{\cal O}_{fb}=\frac{i}{\eta}\int_0^t\, d\tau G(\tau) 
I(t-\tau)\lambda\left[Y_1-Y_2,{\cal O}\right]\,,
\end{equation}
where $G$ is the feedback response function \cite{wismil93}.
Practically, it accounts for the feedback loop delay time, however 
hereafter we assume it negligibly small, hence $G$ can 
be considered as a Dirac delta function.

When inserting Eq.(\ref{Ofb}) into Eqs.(\ref{lang1})-(\ref{lang4}), 
one has to also account for the noise carried inside the system by 
the current (\ref{It}). 
In presence of imperfect detection the current may be non 
trivially related with the input noise.
The following correlations hold at equal times \cite{vit}
\begin{eqnarray}
\langle {\cal W}_j{\cal X}_k \rangle&=&
\langle {\cal X}_j{\cal W}_k 
\rangle=\frac{\sqrt{\eta}}{2}\delta_{jk}\,,
\label{WXcorr}\\
\langle {\cal W}_j{\cal Y}_k \rangle&=&
-\langle {\cal Y}_j{\cal W}_k 
\rangle=i\frac{\sqrt{\eta}}{2}\delta_{jk}\,,
\quad j,k=1,2\,.
\label{WYcorr}
\end{eqnarray}
It is immediate to see that in the perfect efficiency case ($\eta=1$) 
one can identify the feedback noise with the input noise, while in 
the 
opposite case ($\eta=0$) the two noises are uncorrelated (as it can 
be 
easily expected since in such a case the feedback noise has nothing 
to 
do with the vacuum input noise).

It is worth noting that the feedback action (\ref{Ofb}) does not 
affect the steady state values (\ref{Xss})-(\ref{Yss}).
Then we can rewrite Eqs.(\ref{lang1})-(\ref{lang4}) in presence of 
feedback, for only the steady state fluctuations
\begin{eqnarray}
x_j&=&X_j-\langle X_j\rangle_{ss}\,,
\label{x}\\
y_j&=&Y_j-\langle Y_j\rangle_{ss}\,,
\quad
j=1,2\,.
\label{y}
\end{eqnarray}
Let us introduce the operator vectors 
\begin{eqnarray}\label{vn}
v\equiv\left(
\begin{array}{c}
x_1\\y_1\\x_2\\y_2
\end{array}
\right)\,,
\;\;
n\equiv\left(
\begin{array}{c}
{\cal X}_1+\frac{\lambda}{\sqrt{\eta}}({\cal W}_1-{\cal W}_2)\\{\cal 
Y}_1\\
{\cal X}_2-\frac{\lambda}{\sqrt{\eta}}({\cal W}_1-{\cal W}_2)\\{\cal 
Y}_2\\
\end{array}
\right)\,, 
\end{eqnarray}
then, the feedback modified Langevin equations read, in a compact way,
\begin{equation}\label{v-Mvn}
\dot v=-Mv+n\,,
\end{equation}
where
\begin{equation}\label{M}
M=\left(\begin{array}{cccc}
\frac{1}{2}-\lambda&0&\lambda-\chi&0
\\
0&\frac{1}{2}&0&\chi
\\
\lambda-\chi&0&\frac{1}{2}-\lambda&0
\\
0&\chi&0&\frac{1}{2}
\end{array}\right)\,.
\end{equation}
The symmetric noise correlation matrix is
\begin{eqnarray}\label{N}
N&\equiv&\frac{1}{2}\left(\langle nn^T\rangle+\langle 
nn^T\rangle^T\right)
\nonumber\\
&=&\left(\begin{array}{cccc}
\frac{1}{2}+\lambda+\frac{\lambda^2}{\eta}&0&-\lambda
-\frac{\lambda^2}{\eta}&0
\\
0&\frac{1}{2}&0&0
\\
-\lambda-\frac{\lambda^2}{\eta}&0&\frac{1}{2}+\lambda
+\frac{\lambda^2}{\eta}&0
\\
0&0&0&\frac{1}{2}
\end{array}\right)\,,
\end{eqnarray}
therefore the stationary symmetric correlation matrix 
\begin{eqnarray}\label{Ga}
\Gamma\equiv\frac{1}{2}\left(\langle vv^T\rangle+\langle 
vv^T\rangle^T\right)\,,
\end{eqnarray}
can be derived from the typical relation for Ornstein-Uhlenbeck-like 
processes \cite{HSM}
\begin{equation}\label{O-U}
M\Gamma+\Gamma M^T=N\,.
\end{equation}
For the following purposes, we write $\Gamma$ in terms of its 
$2\times 2$ submatrices
\begin{equation}\label{Gags}
\Gamma=\left(
\begin{array}{cc}
\gamma&\sigma
\\
\sigma^T&\gamma
\end{array}\right)\,,
\end{equation}
where the matrix elements, by virtue of Eq.(\ref{O-U}), result
\begin{eqnarray}
\gamma_{11}&=&\frac{(1/2)-\lambda+\left(1-2\chi\right)
\left(\lambda+\lambda^2/\eta\right)}{1-4\lambda+8\lambda\chi-4\chi^2}\,,
\label{g11}\\
\gamma_{12}&=&\gamma_{21}=0\,,
\label{g12}\\
\gamma_{22}&=&\frac{1}{2}\left(\frac{1}{1-4\chi^2}\right)\,,
\label{g22}
\end{eqnarray}
and
\begin{eqnarray}
\sigma_{11}&=&\frac{\chi-\lambda-\left(1-2\chi\right)
\left(\lambda+\lambda^2/\eta\right)}{1-4\lambda+8\lambda\chi-4\chi^2}\,,
\label{s11}\\
\sigma_{12}&=&\sigma_{21}=0\,,
\label{s12}\\
\sigma_{22}&=&-\frac{\chi}{1-4\chi^2}\,.
\label{s22}
\end{eqnarray}

Since the steady state is Gaussian it is completely characterized by 
the correlation matrix (\ref{Ga}).
Then, its degree of entanglement can be quantyified by means of the 
logarithmic negativity \cite{vidwer02}
\begin{equation}\label{L}
L\equiv\left\{
\begin{array}{ccr}
-\log(2\zeta)&\quad&{\rm if}\zeta<1
\\
0&\quad&{\rm otherwise}
\end{array}
\right.\,,
\end{equation}
with 
\begin{equation}\label{z}
\zeta\equiv\sqrt{\left(\det\gamma-\det\sigma\right)
-\sqrt{\left(\det\gamma-\det\sigma\right)^2-\det\Gamma}}\,.
\end{equation}
We have numerically evaluated the quantity
\begin{equation}\label{Lfb}
L_{fb}\equiv\max_{\lambda\in\mathbb{R}}L\,,
\end{equation}
by using Eqs.(\ref{g11})-(\ref{s22}).
In doing so we have
accounted for the stability condition $M\ge 0$ and for the Heisenberg 
uncertainty condition $\Gamma+(i/2)\Omega\ge 0$ \cite{sim}.
Here $\Omega$ is the standard symplectic form
\begin{equation}\label{Om}
\Omega\equiv\left(
\begin{array}{cc}
\omega&0
\\
0&\omega
\end{array}
\right)\,,
\quad
\omega\equiv\left(
\begin{array}{cc}
0&1
\\
-1&0
\end{array}
\right)\,.
\end{equation}
As matter of fact, not all real values of $\lambda$ are admitted 
since the addition of feedback may lead to  unstable or unphysical 
states.
Thus, the maximization (\ref{Lfb}) is performed on a `valid' range of 
$\lambda$'s values. The extension of such a range clearly enlarges as 
$\eta$ decreases. Anyway, the maximum is always achieved for values 
$-(1/2)<\lambda<0$.

\begin{figure}
\begin{center}
\includegraphics[width=0.4\textwidth]{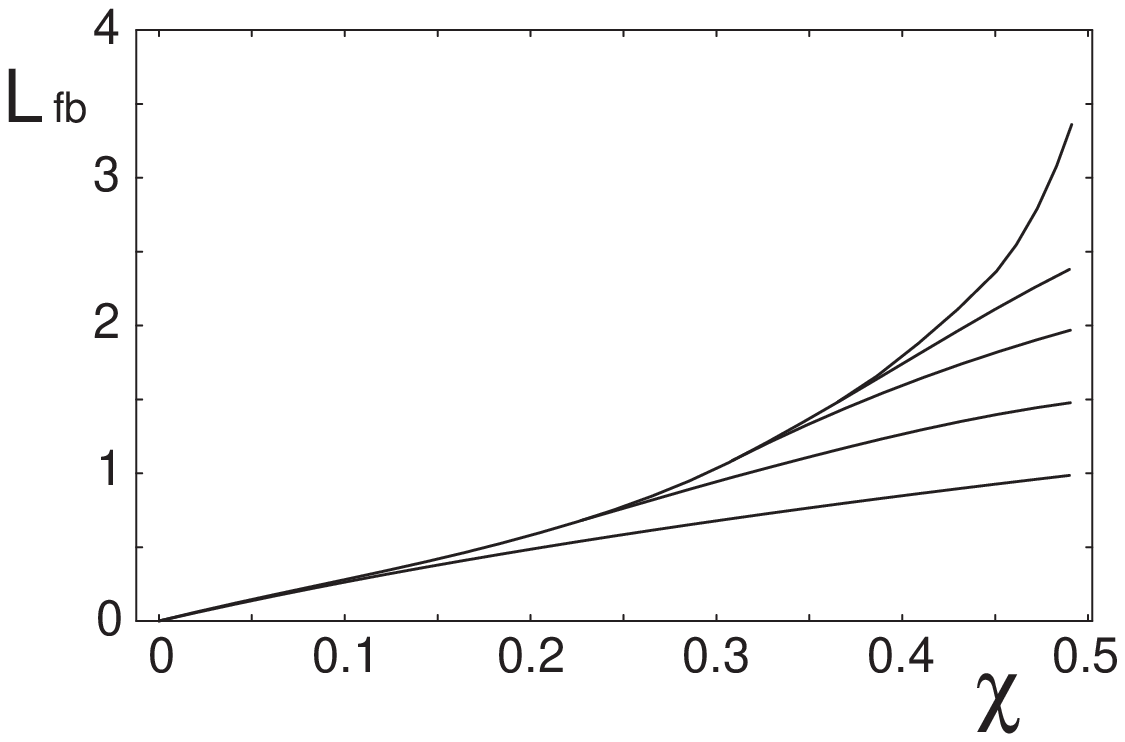}
\end{center}
\caption{\label{fig2} Logarithmic negativity $L_{fb}$ as function of 
$\chi$. 
Curves from bottom to top are for 
$\eta=0$, $0.3$, $0.5$, $0.7$, $0.99$. }
\end{figure}

The quantity $L_{fb}$ of Eq.(\ref{Lfb}) is shown in Fig.\ref{fig2} as 
function of $\chi$ for different values of $\eta$.
Notice that $L_{fb}$ for $\eta=0$ (lower curve) corresponds to the 
case of no feedback.
It tells us that in absence of feedback, entanglement arises for 
$\chi>0$ and icreases 
with it till the value $L=1$. Instead, much higher values can be 
obtained by the feedback action, even with low values of $\eta$. The 
benefit of feedback 
increases by increasing  $\eta$ and it is particularly manifest for 
high values of $\chi$.
In the limit case of $\eta\to 1$ by approaching the instability  
$\chi\to 1/2$ entanglement increases indefinitely.
This is because in such a case the feedback is able to completely 
recycle the information lost by
the system into environment, i.e. it somehow suppresses the amplitude 
damping, making a maximally entangled state achievable  
\cite{reid,epr}.

It is worth noting that 
the presented scheme can be implemented by using an
experimental set up similar to that of Ref.\cite{ouetal},
where the two fields emerging from an optical  parametric oscillator 
were separately subjected to homodyne measurements and the 
relative currents combined.
Then one has to employ driving field modulators to realize the 
feedback action.
Any delays in classical communication (including the feedback loop) 
must be much smaller than the typical time scale of the system 
($\kappa^{-1}$)
and much smaller than the inverse of relevant bandwidth. 
Experimentalists can certainly achieve loop 
delays smaller than $10^{-8}$ s \cite{tauetal} which already satisfies
the above requirements for set ups similar to \cite{ouetal}.
From this point of view Markovian (direct) feedback has also the 
advantage 
of not requiring information processing after measurement.
To give a numerical example, in Ref.\cite{ouetal} an overall 
efficiency of $\sim 0.7$ was achieved, 
which already means the possibility of $\sim 250\%$ improvement of 
entanglement close to the threshold (see Fig.\ref{fig2}).

The feedback action results a Gaussian operation from the quadratic 
form of Eq.(\ref{Hfb}) together with Eq.(\ref{It}). Moreover, the 
presence of vacuum noises ${\cal W}_j$ in such equations
legitimate us to intend the feedback action as local operation 
supplemented by classical communication (LOCC). As matter of fact, in 
absence of interaction ($\chi=0$) the feedback action is unable to 
generate entanglement as it is evident from Fig.\ref{fig2}.
However, in Refs. \cite{eis} it was shown the 
impossibility to enhance (distill) entanglement by means of 
Gaussian LOCC. The key point is that, in contrast with Refs. 
\cite{eis},
we have considered continuous measurement and continuous communication
between the two parties, including the exchange of noise of quantum 
origin.
And the correlations among ${\cal W}_j$ and ${\cal X}_j$, ${\cal 
Y}_j$ are able to reduce the quantum noisy effects on the system, 
hence to enforce the interaction between the subsystems.
This corresponds to an overall nonseparable Gaussian map in the sense 
of  
Ref. \cite{eis}, though the real physical operations are LOCC.
Thus, the presented approach may shed further light on the subject
of entanglement distillation.

It is also to remark that the our model extends behind Eq.(\ref{Hint})
to any other quadratic interaction, because  
Gaussian state's entanglement is always related to the squeezing of a 
suitable quadrature combination
\cite{duan}, which then might play the role of $X_1-X_2$.

In conclusion,
we have shown the possibility to greatly improve the steady 
state entanglement
in an open quantum system by using a feedback action.
In the studied case, where Gaussian states were involved, Markovian 
(direct) 
feedback suffices to reach the goal,
while we guess that other feedback procedures, like
state estimation based feedback \cite{doh99}, 
could be more powerful in other contexts.

\end{document}